\documentstyle[12pt,citesort]{article}
\newcounter{subequation}[equation]

\makeatletter

\def\thesubequation{\theequation\@alph\c@subequation}
\def\@subeqnnum{{\rm (\thesubequation)}}
\def\slabel#1{\@bsphack\if@filesw {\let\thepage\relax
   \xdef\@gtempa{\write\@auxout{\string
      \newlabel{#1}{{\thesubequation}{\thepage}}}}}\@gtempa
   \if@nobreak \ifvmode\nobreak\fi\fi\fi\@esphack}
\def\subeqnarray{\stepcounter{equation}
\let\@currentlabel=\theequation\global\c@subequation\@ne
\global\@eqnswtrue
\global\@eqcnt\z@\tabskip\@centering\let\\=\@subeqncr
$$\halign to \displaywidth\bgroup\@eqnsel\hskip\@centering
  $\displaystyle\tabskip\z@{##}$&\global\@eqcnt\@ne
  \hskip 2\arraycolsep \hfil${##}$\hfil
  &\global\@eqcnt\tw@ \hskip 2\arraycolsep
  $\displaystyle\tabskip\z@{##}$\hfil
   \tabskip\@centering&\llap{##}\tabskip\z@\cr}
\def\endsubeqnarray{\@@subeqncr\egroup
                     $$\global\@ignoretrue}
\def\@subeqncr{{\ifnum0=`}\fi\@ifstar{\global\@eqpen\@M
    \@ysubeqncr}{\global\@eqpen\interdisplaylinepenalty \@ysubeqncr}}
\def\@ysubeqncr{\@ifnextchar [{\@xsubeqncr}{\@xsubeqncr[\z@]}}
\def\@xsubeqncr[#1]{\ifnum0=`{\fi}\@@subeqncr
   \noalign{\penalty\@eqpen\vskip\jot\vskip #1\relax}}
\def\@@subeqncr{\let\@tempa\relax
    \ifcase\@eqcnt \def\@tempa{& & &}\or \def\@tempa{& &}
      \else \def\@tempa{&}\fi
     \@tempa \if@eqnsw\@subeqnnum\refstepcounter{subequation}\fi
     \global\@eqnswtrue\global\@eqcnt\z@\cr}
\let\@ssubeqncr=\@subeqncr
\@namedef{subeqnarray*}{\def\@subeqncr{\nonumber\@ssubeqncr}\subeqnarray}

\@namedef{endsubeqnarray*}{\global\advance\c@equation\m@ne%
                           \nonumber\endsubeqnarray}

\makeatletter
\@addtoreset{equation}{section}
\makeatother
\renewcommand{\theequation}{\thesection.\arabic{equation}}

\def \ci {\cite}
\newcommand{\rf}[1]{(\ref{#1})}
\def \la {\label}
\def \const {{\rm const}}

\catcode`\@=11

\newcount\hour
\newcount\minute
\newtoks\amorpm \hour=\time\divide\hour by 60\minute
=\time{\multiply\hour by 60 \global\advance\minute by-\hour}
\edef\standardtime{{\ifnum\hour<12 \global\amorpm={am}%
        \else\global\amorpm={pm}\advance\hour by-12 \fi
        \ifnum\hour=0 \hour=12 \fi
        \number\hour:\ifnum\minute<10
        0\fi\number\minute\the\amorpm}}
\edef\militarytime{\number\hour:\ifnum\minute<10
0\fi\number\minute}

\def\draftlabel#1{{\@bsphack\if@filesw {\let\thepage\relax
   \xdef\@gtempa{\write\@auxout{\string
      \newlabel{#1}{{\@currentlabel}{\thepage}}}}}\@gtempa
   \if@nobreak \ifvmode\nobreak\fi\fi\fi\@esphack}
        \gdef\@eqnlabel{#1}}
\def\@eqnlabel{}
\def\@vacuum{}
\def\marginnote#1{}
\def\draftmarginnote#1{\marginpar{\raggedright\scriptsize\tt#1}}
\def\draft{
        \pagestyle{plain}
        \overfullrule=2pt
        \oddsidemargin -.5truein
        \def\@oddhead{\sl \phantom{\today\quad\militarytime} \hfil
        \smash{\Large\sl DRAFT} \hfil \today\quad\militarytime}
        \let\@evenhead\@oddhead
        \let\label=\draftlabel
        \let\marginnote=\draftmarginnote
        \def\ps@empty{\let\@mkboth\@gobbletwo
        \def\@oddfoot{\hfil \smash{\Large\sl DRAFT} \hfil}
        \let\@evenfoot\@oddhead}

\def\@eqnnum{(\theequation)\rlap{\kern\marginparsep\tt\@eqnlabel}%
        \global\let\@eqnlabel\@vacuum}  }

\renewcommand{\theequation}{\thesection.\arabic{equation}}
\renewcommand{\thefootnote}{\fnsymbol{footnote}}

\def\appendix#1{
  \addtocounter{section}{1}
  \setcounter{equation}{0}
  \renewcommand{\thesection}{\Alph{section}}
  \section*{Appendix \thesection\protect\indent \parbox[t]{11.15cm}
  {#1} }
  \addcontentsline{toc}{section}{Appendix \thesection\ \ \ #1}
  }
\def \tx {\textstyle}
\def \bi{\bibitem}

\def \ov {\over}
\def \ha {\textstyle { 1\ov 2}}
\def \we { \wedge}
\def \P { \Phi} \def\ep {\epsilon}

\def \go { g_1}\def \gd { g_2}\def \gt { g_3}
\def \gc { g_4}\def \gp {
g_5}
\def \F {{\cal F}}

\def \t {\theta}
\def \p {\phi}
\def \ee {\epsilon}
\def \te {\tilde \epsilon}

\date{}
\begin{document}

\begin{titlepage}

\begin{center}
\hfill UM--TH/00-24\\
\hfill OHSTPY-HEP-T-00-019\\

\vskip 2.5 cm
\vskip 1 cm

{\Large \bf  3-Branes on  Resolved Conifold}

\vskip 1 cm

{\large L.A. Pando Zayas$^a$ and A.A. Tseytlin$^{b,}$\footnote{Also at
 Imperial College, London and Lebedev
Institute, Moscow.}}\\

\end{center}

\vskip 1cm
\centerline{\it ${}^a$ Randall Laboratory of Physics,
The University of Michigan}
\centerline{\it Ann Arbor, MI 48109-1120, USA}
\centerline{\tt lpandoz@umich.edu}

\vskip 0.4 cm

\centerline{\it ${}^b$ Department of Physics, The Ohio State
University,}
\centerline{\it 174 West 18th Avenue, Columbus, OH 43210-1106, USA}
\centerline{\tt tseytlin.1@osu.edu }

\vskip 1.5 cm

\begin{abstract}
The type IIB supergravity solution describing
a collection of regular and  fractional D3 branes
on the conifold (hep-th/0002159) was recently generalized to
the case of the deformed conifold (hep-th/0007191).
Here we present another generalization -- when
the conifold is replaced by the resolved conifold.
This solution
can be found in two different ways:
(i) by first  explicitly constructing
the Ricci-flat  K\"ahler metric on resolved conifold and 
then solving the supergravity 
equations for the D3-brane ansatz with constant dilaton and
(self-dual) 3-form
fluxes; or 
(ii) by generalizing the ``conifold" ansatz  of hep-th/0002159
in a natural
``asymmetric" way so  that the 1-d action describing 
radial evolution
still admits a superpotential, 
 and then solving
the resulting 1-st order system. The superpotentials corresponding 
to the ``resolved" and ``deformed" conifold 
cases turn out to have 
essentially the same simple 
structure.
 The solution in  the resolved conifold case has the same
 asymptotic UV behaviour as in  the conifold case, but unlike
 the deformed conifold case is still singular in the IR.
  The naked singularity is of repulson type and may have
  a brane resolution.

\end{abstract}

\end{titlepage}
\def \K {{\cal K}}
\def\o{\omega}
\def\O{\Omega}
\def\e{\epsilon}
\def\pd{\partial}
\def\pdz{\partial_{\bar{z}}}
\def\bz{\bar{z}}
\def\e{\epsilon}
\def\m{\mu}
\def\n{\nu}
\def\a{\alpha}
\def\b{\beta}
\def\g{\gamma}
\def\G{\Gamma}
\def\d{\delta}
\def\r{\rho}
\def\bx{\bar{x}}
\def\by{\bar{y}}
\def\bm{\bar{m}}
\def\bn{\bar{n}}
\def\s{\sigma}
\def\na{\nabla}
\def\D{\Delta}
\def\l{\lambda}
\def\te{\theta} \def \t {\theta}
\def\ta {\tau}
\def\na{\bigtriangledown}
\def\p{\phi}
\def\L{\Lambda}
\def\hR{\hat R}
\def\ch{{\cal H}}
\def\ep{\epsilon}
\def\bj{\bar{J}}
\def \foot{ \footnote}
\def\be{\begin{equation}}
\def\ee{\end{equation}}
\def \P {\Phi}

\def \ci {\cite}
\def \g {\gamma}
\def \G {\Gamma}
\def \k {\kappa}
\def \l {\lambda}
\def \L {{L}}
\def \Tr {{\rm Tr}}
\def\apr{{A'}}
\def \m {\mu}
\def \n {\nu}
\def \W{{\cal W}}
\def \eps {\epsilon}
\def \ha{{\textstyle { 1 \ov 2}} }
\def \de{{\textstyle { 1 \ov 9}} }
\def \si{{\textstyle { 1 \ov 6}} }
\def \fo{{\textstyle { 1 \ov 4}} }
\def \rt {{\tx { \ta \ov 2}}}

\setcounter{page}{1}
\renewcommand{\thefootnote}{\arabic{footnote}}
\setcounter{footnote}{0}

\def \N{{\cal N}}
\def \ov {\over}

\section{Introduction}

To construct supergravity (and string-theory) duals
of less supersymmetric  gauged theories
one is interested in ``4+1+compact space" type backgrounds with extra
p-form fluxes \ci{pol}.
It is  natural to try  to generalize the original
 AdS/CFT correspondence \cite{ads1} by
considering  D3-branes in more general transverse space
backgrounds, e.g.,
placing them  at conical  singularities \cite{moo,kehagias,kw1,mr}.
 This idea has
been developed further \cite{kg,kn} by adding ``fractional" D3-branes
(D5-branes wrapped over 2-cycles) \ci{gim},
exploiting the fact that topologically  the base space  of the conifold
($T^{1,1}$)
 is $S^2\times S^3$.  In
\cite{kn} it was argued that the dual field  theory should be a
non-conformal ${\N}=1$ supersymmetric  $SU(N+M) \times SU(N)$
gauge   theory  and
some key  effects of introducing fractional branes were
discussed,  including  breaking of conformal invariance and 
structure of the logarithmic RG flow.

 The corresponding 
supergravity solution describing a collection of $N$ D3-branes and
$M$ fractional D3-branes on the conifold
was constructed  in \cite{kts}. This
 solution
 has the standard D3-brane-type metric but with the ``harmonic
function"
 replaced by $h(r) = 1 + {Q(r) \ov r^4}, $ \
$Q(r)= c_1 g_s N  +   c_2 (g_s M)^2 \ln {r\ov r_0} $.
 The logarithmic running of the ``effective charge"
$Q(r)$ implies the presence of a naked singularity at small $r$
 (in IR from dual gauge theory point of view).
 A remarkable way to avoid the IR  singularity  was
  found  \cite{ks}: one is to replace  the conifold
   by the  deformed conifold  keeping the same D3-brane
   structure of the 10-d metric  and generalizing
    the 3-form ansatz  appropriately.
    
The deformed conifold solution \cite{ks}   has
the same large $r$  (UV) asymptotic as the original
 conifold one  \ci{kts}   but is regular at small $r$, 
 i.e. in the
 infrared. 
In \cite{ks} some  desirable properties of this background were
established, including the gravity counterparts 
of the  existence of confinement and chiral
symmetry breaking in the dual gauge theory.

It is of obvious interest to  explore further
 the class of  backgrounds which
have similar ``3-branes on conifold" type  structure
(potentially including also the solution of 
\ci{MN}).
Given the topology of the base space, there are two natural ways of
smoothing out the singularity at the apex of the conifold. One can
substitute the apex by an $S^3$ (deformation) or by an $S^2$
(resolution). Here we complement the
conifold  \ci{kts} and deformed conifold  \ci{ks}  
 solutions by constructing explicitly the background
  corresponding to
the
  resolved conifold case.\foot{
  Various aspects of branes on resolved
  conifold
  were discussed, e.g.,  in \ci{vaf1,kw2,vaf2}.}
  The type IIB supergravity
  solution  we find
coincides with the original   background of \ci{kts}
 for large $r$ 
 but has  somewhat  different (though still singular)
 small $r$ (IR) behavior.
  The singularity of the  analog of the 
 solutions of  \ci{kts,ks}  in the resolved
 conifold case was anticipated in \ci{ig,polg}.

One may  discover this solution using two different
strategies. One may start with the ``conifold" 
 ansatz for the 10-d background 
 in  \ci{kts} and generalize it 
in a very simple and natural
way  by allowing an ``asymmetry" between the two $S^2$ parts
 in the  metric and in the NS-NS 3-form  introducing two
 new functions. Assuming the spherical symmetry 
  as in \ci{kts} one can then  obtain
  the resulting supergravity equations from a 1-d action
  describing evolution in radial direction.
 Remarkably, as in \ci{kts}, 
 the potential term in this action
 can be derived from a superpotential. This 
 is true  also in  the ``deformed" case  of \ci{ks} 
  and is consistent with expected $\N=1$ supersymmetry
  of the resulting solution which follows then
   by solving the resulting
  1-st order equations. In the process,  one 
  explicitly determines
  the metric on the resolved conifold.

Alternatively, one may start with finding  
 the Ricci flat  K\" ahler metric of the resolved conifold
(which, as far as we know,
 was not previously given   in the literature in an explicit form)
  and then solve the type IIB supergravity 
equations for the D3-brane ansatz with constant dilaton and  3-form
fluxes representing the inclusion of fractional D3-branes.
As in the other two (``standard" and ``deformed"
conifold) cases, the complex 3-form
field turns out to be self-dual. The importance of this
 property was emphasized in \ci{ks,ig} and
 the  $\N=1$ supersymmetry
 of such  class of backgrounds was recently proved  
 in \ci{polg,gub}.

In section 2 we shall
 review the ``small resolution" of the conifold \cite{candelas}
 and find the corresponding
Ricci flat  metric explicitly.

In section 3 we shall show how the geometry of  $M^{10} = R^{1,3}
 \times$(resolved conifold) changes in the presence of
  D3-branes, i.e find the analog  of the standard
  D3-brane solution \ci{horst,dul}
  in case when the transverse 6-space is replaced by
  the resolved conifold (the  coefficient in
  the metric is  a harmonic function on the 6-space).
In contrast to the D3-brane on the conifold  \ci{kw1}  the
short distance limit of this supergravity
background   does not have an $AdS_5$ factor and is singular.
We shall compare this D3-brane  solution
with the one  in the case of the deformed conifold \ci{ks}.
We
consider a radially symmetric solution corresponding  to
3-branes smeared over a 2-sphere at the apex. In \ci{kw2} the 
3-branes were instead localized at
a point on $S^2$. The choice of  point corresponds to giving
expectation
values for scalar fields in the dual field theory, 
breaking gauge symmetry and conformal invariance. 
The solution in \ci{kw2} was  non-singular in IR, approaching 
$AdS_5 \times  S^5$. It seems that the averaging 
over $S^2$ causes a singularity (present also in  analogous 
D3 brane solution  with 
 deformed conifold as transverse space).\foot{We 
 are grateful to I. Klebanov for this suggestion 
 and explaining the relation to  \ci{kw2}.}

 In section 4 we shall generalize the D3-brane solution  of Section 3
 to the presence of  fractional D3-branes
   on the resolved conifold. We  shall
   analyze the limits
   of the solution and show that it has a short-distance singularity.
   This is
a repulson-type  singularity, so  one may hope
that it may be resolved by  the  mechanism of
 \cite{enh1}.


 In Section 5 we  shall explain how the same solution can be obtained
 from a 1-d action for radial evolution
  admitting a superpotential,
 i.e. by solving a system of 1-st order equations as in \ci{kts}.
 We shall point out    that a similar superpotential exists also in
 the deformed conifold case  of \ci{ks}. As we  shall demonstrate
 in the process, 
  making simple ansatze 
  for the 6-d part of the metric  and identifying  the
    1-st order systems
associated with the 
Ricci-flatness equations  allows one to find the explicit 
forms of 
 the resolved  and deformed  conifold metrics
in a straightforward way.
 Identifying explicitly 
 the  1-st order system (whose existence is
 expected on the grounds of residual 
 supersymmetry) is useful for generalizations 
 and for  establishing a potential
  correspondence 
 between the radial evolution on the supergravity side and
 the ${\cal N}=1$ supersymmetric
 RG  flow in the dual gauge theory.


\section{Metric of resolved conifold  }

The  purpose of this section is to  write down explicitly
the Ricci flat  metric
on the resolved  conifold following the
detailed discussion in  \cite{candelas}.
 Though  it is not possible to introduce
  a globally well-defined metric
 on the resolved conifold, one
can  find  a metric on  each of the two covering
patches.
The conifold can be described by the following quadric in ${\bf
C}^4$:
$
\sum_{i=1}^4 w_i^2=0.\foot{More details on the
topological structure of the resolved conifold as a ${\bf C}^2$
bundle
over ${\bf CP}^1$ can be found in \cite{candelas}.}
$
This equation can be written as
\begin{equation}\mbox{det}\ \W=0\ , \ \ \ \  
 {\rm i.e.} \ \ \ \ XY-UV=0\
,
\ee \be
\W=
{1\over \sqrt{2}}\left(
\begin{array}{cc}
w_3+iw_4&w_1-iw_2\\
w_1+iw_2&-w_3+iw_4
\end{array}
\right)
\equiv
\left(
\begin{array}{cc}
X&U\\
V&Y
\end{array}
\right).
\end{equation}
 The resolution of
the conifold can be  naturally  described 
in terms of $(X,Y,U,V)$.
Resolving the conifold means  replacing  the equation
 $ XY-UV=0$  by the  pair of equations
\begin{equation}
\label{rconifold}
\left(
\begin{array}{cc}
X&U\\
V&Y
\end{array}
\right)
\left(
\begin{array}{c}
\l_1\\
\l_2
\end{array}
\right)
=0\ ,
\end{equation}
where $\l_1 \l_2 \not=0$.
Note that $(\l_1,\l_2)\in {\bf
CP}^1$  (any pair
obtained from a given one
 by multiplication by a nonzero complex number is also a solution).
  Thus $(\l_1,\l_2)$
is uniquely  characterized by  $\l=\l_2/\l_1$
in the region where $\l_1\ne 0$. Working on this patch a solution to
(\ref{rconifold}) takes the form\footnote{In the region
where $\l_1$ is allowed to be zero we have  $\l_2\ne 0$ and thus the
general solution can be written as $\W=
\left(
\begin{array}{cc}
X&-X\m\\
V&-V\m
\end{array}
\right), 
$ where $\m=\l_1/\l_2$.}
\begin{equation}
\label{rw}
\W=
\left(
\begin{array}{cc}
-U\l&U\\
-Y\l&Y
\end{array}
\right) \ .
\end{equation}
 Thus $(U,Y,\l)$ are the three complex coodinates characterizing
the resolved conifold in the patch where $\l_1\ne 0$.

The conifold metric is
  $g_{m{\bar n}}=\pd_m\pd_{\bar
n}{\rm K}$,  where ${\rm K}$ is the K\"ahler potential.
In contrast to the the cases of
the conifold or the deformed conifold,
here   the
K\"ahler potential is not a globally defined quantity, and
is not
 a function of only the  radial coordinate defined by
 \begin{equation}
r^2=\mbox{tr} (\W^{\dagger} \W) =(1+|\l|^2)(|U|^2+|Y|^2)\ .
\end{equation}
 Following the analysis
of \cite{candelas}, based on the transformation of the coordinates in
the overlap region, one concludes that the most general K\"ahler
potential is of the form \be
{\rm K}=F(r^2) +4a^2\ln (1+|\l|^2)\ , \ee
  where $F$ is a
function of $r^2$ and $a$ is the ``resolution" parameter
($a=0$ is the conifold case).
Thus the metric is
\begin{equation}
\label{crcon}
ds^2=F'\mbox{tr}(d\W^{\dagger}d\W)+F''|\mbox{tr}(\W^{\dagger}d\W)|^2 +
4a^2{|d\l|^2\ov (1+|\l|^2)^2}\ ,\ \ \ \ \  F'\equiv { dF\ov dr^2} \ .
\end{equation}
 The Ricci
tensor for a K\"ahler metric is $R_{m{\bar n}}=-\pd_m\pd_{\bar n}\ln
\mbox{det}g_{m{\bar n}}$, where  for the metric in
(\ref{crcon})
\begin{equation}
\mbox{det}g_{m{\bar n}}=F'(F'+r^2F'')(4a^2+r^2F')\ .
\end{equation}
The Ricci-flatness condition  implies
\begin{equation}
\label{gam1}
\g'  \g(\g + 4a^2)={2\over 3}r^2\ ,\ \ \ \ \ \ \ \ \ 
\g\equiv r^2 F' \ ,\ \  \ \ \  \g'\equiv  {d\g\ov d r^2}\ ,
\end{equation}
 which is integrated to give
\begin{equation}
\label{gam2}
\g^3+6a^2\g^2-r^4=0\ .
\end{equation}
We set the integration constant to zero, assuming that
$\g(0) =0$ (as should be
 true  in the $a=0$  case  of  the conifold).
The  real solution is\foot{This 
expression applies  for all $r^2 >0$ 
 provided  
for  $r^2<  4 \sqrt{2} a^3$  one uses the cubic root
$(-1)^{1/3} = { 1 + i \sqrt 3 \ov 2}$   
 (while $N$ becomes complex, 
$\gamma$ stays real).}
\begin{equation}
\label{gamma}
\g=-2a^2+4a^4 N^{-1/3} (r)+{N^{1/3}(r)}, \qquad \ \ \
N(r)\equiv \ha (r^4-16a^6 +\sqrt{r^8 -32a^6r^4})\ .
\end{equation}
In the conifold case  $a=0$ we have $\g=r^{4/3}$
\cite{candelas}. Note also  that
\begin{eqnarray}
\label{asymptotics}
\g(r\to 0)={1\over \sqrt{6} a}r^2 -{1\over 72 a^4}r^4+
{O}(r^6)\ , \qquad \ \
\g(r\to\infty)=  r^{4/3}-2a^2+ {O}(r^{-4/3})\ .
\end{eqnarray}
To write down the resolved conifold metric
explicitly we will parametrize $\W$
in terms of the two sets of 
 Euler angles, exploiting the fact that the
 resolved conifold solution for $\W$ has
  $SU(2)\times SU(2)$ symmetry\foot{Here
  $\psi= \psi_1+ \psi_2$, and
  $(\te_1,\p_1,\psi_1)$  and $(\te_2,\p_2,\psi_2)$
  correspond to the two $SU(2)$'s.}
\be
U=r e^{{i\over
2}(\psi+\p_1+\p_2)} \cos{\tx{ \te_1\over 2}}\cos{\tx {\te_2\over 2}}\ ,
\ \ \
Y= re^{{i\over
2}(\psi-\p_1+\p_2)}\sin{\tx{ \te_1\over 2}}\cos{\tx {\te_2\over 2}}\  ,
\ \ \
\l= e^{-i\p_2}\tan{\tx {\te_2\over 2}}.
\ee
Then the  resolved conifold metric  takes the form
$$
ds^2_6 =\g'dr^2+\fo {\g}\sum_{i=1}^2 \left(d\te_i^2+\sin^2\te_i
d\p_i^2\right) + \fo {\g' r^2}\big(d\psi + \sum_{i=1}^2 \cos\te_i
d\p_i\big)^2 $$
\be +\ a^2(d\te_2^2+\sin^2\te_2 d\p_2^2)\ .
\end{equation}
Note that the parameter $a$ introduces asymmetry between the two
 spheres.
Defining  the veilbeins
\begin{equation}
 e_{\psi} = d\psi + \sum_{i=1}^2 \cos \te_i d\p_i \ ,
\qquad e_{\te_i} = d\te_i\ ,
\qquad e_{\p_i} = \sin \te_i d\p_i \ , \ \ \ \ \  i=1,2\ ,
\end{equation}
the metric  can be written as
\begin{equation}
ds_6^2=\g'dr^2 +\fo{\g' r^2} e_{\psi}^2
+\fo{\g
}\left(e_{\te_1}^2+e_{\p_1}^2\right)+ \fo  ({\g} +4
a^2)\left(e_{\te_2}^2+e_{\p_2}^2\right)
 \ .
\end{equation}
As follows from (\ref{asymptotics}),
 for small $r$ the
$S^3$ ($\psi,\theta_1,\phi_1$)  part of the metric
 shrinks to zero size while the $S^2$ $(\te_2,\p_2)$ part
stays finite with  radius $a$.

Since $\g'd(r^2) =d\g$  and $\g'= {2 r^2\ov 3 \g (\g + 4 a^2)}$
it is very 
  convenient to  consider $\g$ as a new radial
coordinate 
 introducing 
\be \rho^2\equiv {3\ov 2} \gamma\  \ee
(since $0 \leq r <  \infty$, we also have $0\leq \rho <  \infty$).
This allows one to avoid the  issue of how to define
the expression \rf{gamma} in different regions.
Then
using (\ref{gam1}),(\ref{gam2}) the resolved
conifold metric can be written simply  as
\begin{equation}
ds_6^2={ \k^{-1}(\r)}d\r^2+ \de  \k (\r) \r^2 e_{\psi}^2 +
\si \r^2\left(e_{\te_1}^2+e_{\p_1}^2\right)+\si
 ({\r^2} +
6a^2)\left(e_{\te_2}^2+e_{\p_2}^2\right)\ , \la{reso}
\end{equation}
where
\begin{equation}
\k(\r)\equiv {\r^2+9a^2\over \r^2+6a^2}\ .
\end{equation}
This is the explicit $SU(2)\times SU(2) $
invariant
form of the resolved conifold   metric  which
we shall use in what follows.\foot{It is easy to check directly that
this
metric is indeed Ricci flat.}
 When the resolution parameter $a$ 
goes to zero  or  when  $\rho\to \infty$
 it reduces to the standard
conifold metric with $T^{1,1}= {SU(2) \times SU(2) \ov U(1) }$
as the  base \ci{candelas,page}
\begin{equation}
(ds_6^2)_{\r\to \infty}  =d\r^2+ \r^2 \bigg[ \de  e_{\psi}^2 +
\si \left(e_{\te_1}^2+e_{\p_1}^2 +
e_{\te_2}^2+e_{\p_2}^2\right)\bigg] \ .
\end{equation}
For small $\r$ the metric \rf{reso} reduces to
\begin{equation}
(ds_6^2)_{\r\to 0} ={\textstyle  { 2 \ov 3}}  d\r^2+ \si
 \r^2 \left( e_{\psi}^2 +
e_{\te_1}^2+e_{\p_1}^2\right)+
( a^2  + \si \r^2)    \left(e_{\te_2}^2+e_{\p_2}^2\right) \ .\la{sma}
\end{equation}
This shows once again
 that near the  apex $(\r=0)$ the $S^3$ part shrinks to zero size
while the radius of $S^2$ $(\te_2, \p_2)$  part 
 approaches finite value equal to $a$.
This  metric with $a\not=0$ 
 has a regular curvature  --
in contrast to the standard conifold  here the curvature invariants 
are regular at $\r\to 0$ with $|R_{....}|_{\r\to 0}  \sim   { 1 \ov a^2}$, 
 i.e. the parameter $a$ plays indeed the role of the  
singularity resolution parameter.


\section{D3-branes on  resolved conifold}

As is well known, given a Ricci flat 6-d space with
 the metric $g_{mn}$
one can construct the following generalization
of the standard \cite{horst,dul}
 brane solution (see, e.g.,
\cite{kehagias,minasian,papa})
\be
ds^2_{10}=h^{-1/2}(y)dx^{\m}dx^{\m}+h^{1/2}(y)g_{mn}(y)dy^{m}dy^{n}\ ,
\la{meet}\ee
\be
F_5=(1+*)d h^{-1}\wedge dx^0\wedge dx^1\wedge dx^2\wedge dx^3 \ ,
\ \ \ \ \ \ \Phi=\const\ , \la{typ}
\ee
where  $h$
is a  harmonic function  on the transverse 6-d space:
\begin{equation}
\label{d3eqn}
{1\over \sqrt{g}}\pd_m\left(\sqrt{g}g^{mn}\pd_n h\right)= 0\ .\la{har}
\end{equation}
 Let us solve \rf{har} for the resolved conifold metric
 (\ref{reso})
  assuming
  $h=h(\r)$. Using that
$\sqrt{g}=
{1 \ov 108} \r^3(\r^2+6a^2)\sin\te_1\sin\te_2$  we get
\begin{equation}
h=h_0 +{2{L}^4\over 9a^2 \r^2}-{2{L}^4 \over 81 a^4}\ln(1+{9a^2\over
\r^2})\ ,
\end{equation}
where we have chosen the integration constant so that in the
$a\to 0$ (or, equivalently, large $\r$) limit
the solution approaches the
standard flat space or conifold one
\be
h(\rho \to \infty) =h_0+{{L}^4\over \r^4}\ .
\ee
For small values of the radius $\r$  we get\footnote{
It is easy to check  that $h$ does not vanish at any real value of $r$.
Introducing $x={9a^2\ov\r^2}$ and $c^2={81a^4h_0 \ov 2{L}^4}$
the equation $h(r)=0$
becomes
 $\ln(1+x)-x =c^2$ which has no $x > 0$ solutions.}
\be
h(\r \to 0) = {b^2\over  \r^2}\ , \ \ \ \ \ \   \  \ \ \ \ \
b^2={ 2 {L}^4 \ov  9 a^2}\ ,
\ee
so that the  10-d metric becomes
$$
(ds^2_{10})_{\r\to 0}
={\r\over b} dx^\m dx^\m+ {b\over \r} (ds_6^2)_{\r\to 0}\ $$
\be
=\  b \bigg[  {y^2 \over b^2} dx^\m dx^\m    +
           {\textstyle  { 8 \ov 3}}  dy^2+ \si  y^2 \left( e_{\psi}^2 +
e_{\te_1}^2+e_{\p_1}^2\right)+
{ a^2\ov y^2} \left(e_{\te_2}^2+e_{\p_2}^2\right) \bigg]
 \ , \la{smae}
\end{equation}
where $y=\sqrt {\rho}$ and  we used the expression \rf{sma}
for the short-distance  limit
of the
resolved conifold metric.

The $S^3$ part of the 10-d metric
still shrinks to zero size but the
size of $S^2$ rather than approaching the constant  $a$
now blows up
at $\r=0$.  It is easy to check that
the point $\r=0$ is the curvature singularity
(while the Ricci scalar vanishes as
for any D3-brane
solution of the type \rf{typ}, the Ricci tensor is singular).
This  behaviour
is to be compared
with one in the case of the D3-branes at  the conifold singularity
where the  short-distance limit of the geometry  was regular
$AdS_5 \times T^{1,1}$  space
(see \cite{kw1}).


For completeness, let us compare the above solution with the one
in the case  when the transverse 6-space is the deformed conifold with
the metric
   \cite{candelas,minasian,ohta,ks}
\begin{equation}\la{ddd}
ds^2_6
=  \ha \ep^{4/3} \K
\left[
(3\K^3)^{-1}(d\ta^2+g_5^2)
+
{\rm sinh}^2{\tx {\ta\over 2}}\ (g_1^2+g_2^2)
+{\rm cosh}^2{\tx {\ta\over 2}}\ (g_3^2+g_4^2)\right],
\end{equation}
where $ \ep$ is the deformation parameter, 
\begin{equation}\la{kee}
\K(\ta)={[\ha \sinh(2\ta)- \ta]^{1/3}\over \sinh\ta} \ ,
\end{equation}
and the 1-forms $g_n$ defined in \ci{ks} are 
\be\la{gee}
g_1 = - {\eps_2 +  e_{\p_1} \over\sqrt 2}\  ,\ \ \ 
g_2 = - {\eps_1  - e_{\t_1} \over\sqrt 2}\ , \ \ \ 
g_3 = {\eps_2 -  e_{\p_1}\over\sqrt 2}\  ,\  \  \
g_4 = {\eps_1 + e_{\t_1}  \over\sqrt 2}\ ,\ \ \ 
g_5 = e_\psi\ ,
\ee
$$
 \eps_1\equiv \sin\psi\sin\theta_2 d\phi_2+\cos\psi d\theta_2
\  , \ \ \ \ \  
\eps_2
\equiv   \cos\psi\sin\theta_2 d\phi_2-\sin\psi d\theta_2\ . $$
The harmonic function $h$ in \rf{meet} 
  is then found to be
\begin{equation}
h(\tau)=h_0 -  h_1 \int {d\ta\over [\ha \sinh
(2\ta)-\ta]^{2/3}}
\ =\ \left\{
\begin{array}{cc}
 1+  { 3 \ov 4^{2/3}}  h_1{ e^{-4\ta/3}}, & \ \ \ta \to \infty\\
\,&\, \\
({ 3 \ov 4})^{2/3} h_1 \ta^{-1}, & \ \ \ta \to 0 \\
\end{array}
\right.
\end{equation}
Introducing $\r \sim e^{\ta/3}$ for large $\tau$ we recover
 the D3-brane on the conifold limit with $h= h_0 + { {L}^4\ov \r^4}$.
 For small values of $\tau$  we get
\begin{equation}
ds_{10}^2={\sqrt{\r}\over m}\eta_{\m\n}dx^\m dx^\n + {m\over
\sqrt{\r}} (ds_6^2)_{\r\to 0} \ , \ \ \
(ds_6^2)_{\r\to 0}=
 d\r^2+\ha {\r^2
 }d\O_2^2 +{\ep^{4/3}\over (12)^{1/3}}d\O_3^2 \ , \ee
where $
 \r\equiv
{\ep^{2/3}\over 2^{5/6}3^{1/6}}\ta$,
and   $m^2=  {L}^4  \ep^{-2}(3/2)^{-5/2}$.
In the short distance  limit of the  6-d
deformed conifold metric
the  2-sphere
shrinks to zero size  while the 3-sphere  part has finite radius
related  to the deformation parameter $\eps$.
In the 10-d metric
 the $S^2$ part still shrinks to zero size but the radius of the
  $S^3$ part
blows up at the point  $\r=0$  which is the curvature singularity.
 As in the resolved conifold case,
  the near-core geometry is singular.
That is why to get a regular solution after adding
 fractional D3-branes \ci{ks}
one  needs to set  the ``bare"  D3-brane
charge to zero to make possible for the 5-form field 
 (and thus for the Ricci tensor)  to vanish at small $\rho$.


\section{Fractional D3-branes on  resolved conifold}

Let us now study a generalization of the D3-brane  solution
of the previous section to the case  of additional 3-form fluxes,
with the aim to find the  analog  of the solution
of \ci{kts} describing a collection of regular and fractional
D3-branes on the conifold  in  the case when the conifold is 
replaced by
the resolved conifold. 
 The ansatz for the  metric will  be the same as in \rf{meet},
 \be
 ds_{10}^2=h^{-1/2}(\r)dx^\m dx^\m +h^{1/2}(\r)ds_6^2\ , \la{mee}
\end{equation}
where $ds_6^2$  will be  the metric of the resolved conifold
(\ref{reso}).
Our ansatz for the NS-NS
2-form  will be  a natural  generalization of the ansatz
in \ci{kts}  motivated by an asymmetry between the two $S^2$
parts in the resolved conifold metric \rf{reso}
$$
B_2 = f_1(\r)e_{\te_1}\wedge
e_{\p_1}+f_2(\r)e_{\te_2}\wedge e_{\p_2} \ , $$
\be
\label{fields}
H_{3}=dB_2= d\r\wedge[f'_1(\r)e_{\te_1}\wedge
e_{\p_1}+f'_2(\r)e_{\te_2}\wedge e_{\p_2}]\ .
\ee
The ``conifold"   ansatz \ci{kn,kts} corresponds to
$f_1=-f_2 $. The ansatz for the R-R 3-form $F_3 $ is 
dictated by the  closure  condition
 $dF_3=0$, i.e. the forms $F_3$ and $F_5$
 will be  taken in  the same form as in \ci{kn,kts}\foot{Note that
 our definition of the  basis of  1-forms differ from \ci{kts}
 by numerical factors, so that  the constant $P$  and function $K$
 here  are  related to the ones in
  \ci{kts} by $P\to  { 1 \ov 18  \sqrt 2 } P$,
  \  $K\to  { 1 \ov  108} K.$
  Also, in the case of \ci{kts} $f_1=-f_2 = { 1 \ov 6\sqrt 2 } T$.}
\be \la{rrr}
F_3 = P e_{\psi}\wedge (e_{\te_2}\wedge e_{\p_2}- e_{\te_1}\wedge
e_{\p_1} )\ , \ee \be \la{fre}
F_5=  {\cal F}+* {\cal F}\ , \quad  \ \ \ \
{\cal F} = K(\r)e_{\psi}\wedge e_{\te_1} \wedge e_{\p_1} \wedge
e_{\te_2}\wedge e_{\p_2}\ .
\ee
Then  using the metric \rf{reso}
the 10-d duals of these  forms are  found to be
\be
{*}{\cal F}={108  K \over\r^3 (\r^2+9a^2) h^2 }d\r\wedge
dx^0\wedge dx^1 \wedge dx^2 \wedge dx^3\ ,
\ee
\be
{*} F_3={3P\r\over (\r^2+9a^2) h  }d\r\wedge dx^0\wedge dx^1 \wedge
dx^2 \wedge dx^3\wedge \left( e_{\te_1}\wedge e_{\p_1}-\G^2 e_{\te_2}
\wedge e_{\p_2} \right)\ , \ee \be
{*} H_{3}=-{\r^2+9a^2\over 3 \r h }dx^0\wedge dx^1 \wedge dx^2 \wedge
dx^3\wedge e_{\psi} \bigg(f'_1 e_{\te_2} \wedge e_{\p_2}
 + \G^{-2}{f'_2}e_{\te_1}\wedge e_{\p_1} \bigg)\ . \label{hodgedual}
\ee
Here
\be \G \equiv {\r^2   + 6a^2\ov \r^2 } \ . \ee
 is  the ratio of the squares of the  radii
of the two spheres in the metric \rf{reso} and its difference 
from 1 is 
 a signature of the resolution ($a\not=0$).

 As in \ci{kts,ks} we shall assume that the dilaton $\Phi$ is constant.
 Then the  $F_3$ equation
of motion
$
d(e^\Phi *F_3)=F_5\wedge H_{3}
$
is satisfied automatically, and from the $H_{3}$ equation
$
d(e^{-\Phi} * H_{3})=-F_5\wedge F_3
$
one obtains  the following three   equations ($ e^\Phi= g_s$)
\be
\bigg[{f'_1(\r^2+9a^2)\over h\r}\bigg]'=
{324g_s P
K\over h^2\r^3(\r^2+9a^2)}\ , \ \ \ \
\bigg[{f'_2(\r^2+9a^2)\over h\r\G^2}\bigg]'=-{324g_s P
K\over h^2\r^3(\r^2+9a^2)}\ , \ee \be
f'_1+ \G^{-2}f'_2=0 \ . \la{las}
\ee
It follows 
from \rf{las} that for $\G=1$
one should have $f_2=-f_1$
(modulo an
irrelevant constant)
which  was   precisely the assumption of \cite{kts} in the $a=0$ case.

The constant dilaton
 condition implies $H_{3}^2=e^{2\Phi} F_3^2$, i.e.
 using \rf{reso} we get\foot{As in
 \ci{kts,ks}, the axion
 equation is satisfied automatically since $H_3\cdot F_3 =0$.}
\begin{equation}
f'^2_1+ \G^{-2}f'^2_2={9 g_s^2P^2\over k^2\r^2}\left(1+
\G^{-2} \right)\ .
\end{equation}
Combined with \rf{las} that gives
\be
f'_1= 3g_sP {\r \over \r^2+9a^2}\ , \qquad \quad
f'_2=-3g_sP{(\r^2+6a^2)^2 \over \r^3(\r^2+9a^2)}\ .
\la{fff}
\ee
It is easy to see from the above relations
that, as in the conifold \ci{kts} and the  deformed conifold
cases \ci{ks},
the forms $H_3$  and $F_3$ are  dual to each other
 in the 6-d sense.
This property, together with the Calabi-Yau  nature of the
(original, deformed or resolved)
conifold metrics  implies the $\N=1, d=4$ supersymmetry of the
resulting  backgrounds \ci{polg,gub}.

 The Bianchi identity for the 5-form \ $
d*F_5=dF_5=H_{3}\wedge F_3$ gives
\begin{equation}
K'= P (f'_1-f'_2)\  , \ \ \  \  {\rm i.e. } \ \
 \ \ \   K   = Q + P (f_1-f_2)  \ . \la{kkk}
\end{equation}
The symmetries of the
 metric ansatz imply (again, as in the other two conifold
 cases\ci{kts,ks})
 that to determine the function  $h(\r)$  it is sufficient
 to consider the
trace of the Einstein equations,  
$R = - \ha \Delta h = { 1 \ov 24}
( e^{-\Phi} H_3^2 + e^{\Phi} F_3^2) $, i.e. 
\begin{equation}
h^{-3/2}{1\over \sqrt{g}}\pd_{\r}\left(\sqrt{g}g^{\r\r}\pd_{\r} h
\right)
= - {\tx { 1 \ov 12}}  (g_s^{-1} H_{3}^2+g_s F_3^2) = -\si  g_sF_3^2\ ,
\end{equation}
where $g_{mn}$ is the 6-d metric \rf{reso}
($g^{\r\r}= \k(\r)= { \r^2+9a^2\ov \r^2+6a^2}
, \ \sqrt{g}\sim  \r^3(\r^2+6a^2)$), i.e.
 \begin{equation}
\left[\r^3(\r^2+9a^2) h' \right]'=
- 324 g_sP^2 { \r ( 1+\G^2)
\over \r^2+9a^2}\  .
\end{equation}
Integrating this equation we get
\be
h'  =-  { 36 g_s P^2 \ov \r^3 ( \r^2 + 9 a^2)}
 \left(  3 Q  -   { 18 a^2 \ov \r^2}  +  \ln[ \r^{8}(\r^2 + 9a^2)^5]
\right)
\ ,  \la{pio}
\ee
where we have chosen the integration constant to be related to the
one in \rf{kkk}. From  \rf{fff} we find (we omit trivial constants of integration)
\be
f_1(\r) =
{\tx {3\over 2}}  g_s P\ln (\r^2+9a^2)\ , \ee
\be
f_2(\r) =   \si g_s P \bigg(  {36a^2\over \r^2}-
 \ln [\r^{16} (\r^2+9a^2)] \bigg)\ , \ee
and thus from \rf{kkk}
\be
K(\r) = Q  - {\tx {1 \ov 3}}  g_s P^2  \bigg(  {18a^2\over \r^2}-
 \ln [\r^{8} (\r^2+9a^2)^5] \bigg)\ . \la{kik} \ee
Note that \rf{pio},\rf{kik} imply  that
\be \la{kiik}
h'= -{108 \r^{-3}(\r^2+9a^2)^{-1} K(\r) }\ . 
\ee
Integrating \rf{pio} one can  find  the explicit
form of $h(\r)$  which is not very illuminating as it contains
the special function  $Li_2(-{\r\ov 3a})$.
The constants $Q$ and $P$ are proportional to the numbers
 $N$ and $M$ of
regular and fractional D3-branes.

In the large $\r$ $({\r \gg 3 a })$ limit we reproduce the
solution of \cite{kts} with its characteristic logarithmic behavior
\be
f'_1 = 3g_sP{\r^{-1} }\ ,
 \qquad f'_2=-3g_sP{\r^{-1}}\ ,
\ \ \ \  K= Q + 6g_sP^2\ln \r \   ,
\ee
\be
h= h_0 + { L^4
  + 162g_s P^2 (\ln \r  + \fo )  \over \r^4}\ , \
\ee
where $  {L}^4 = 27 Q   $
(and $h_0 = g^{-1}_s$ as we use the Einstein-frame metric).

 In the short distance limit $(\r\ll 3a )$ 
  the solution becomes
\be
f'_1={g_sP\over 3a^2}\r\ , \ \ \qquad f'_2=- {12g_s Pa^2\ov \r^3}
 \ ,\ \ \ \ \ \ 
K= Q -{6g_sP^2a^2\over \r^2}\ , \ee
\be
h= h_0+{6Q\over a^2\r^2}-{18 g_s P^2\over \r^4}\ .
\ee
The form of $h$ implies the presence of  a naked
singularity  at $\r=\r_h$ 
\begin{equation}
\label{naked}
\r_h^2= {3Q\ov h_0a^2}(\sqrt{1+2h_0g_sP^2a^4Q^{-2}}-1)\ .
\end{equation}
For small number of fractional D3-branes $(P \ll Q)
$ the singularity is located at
$\r^2_h= 3g_sP^2 Q^{-1} a^2$. \footnote{One obtains the same value
 by simply  sending $h_0\to 0$, as naively expected 
  in the limit of small radius.}
At the same time,   the five-form  coefficient $K(\r)$ \rf{kik}
  vanishes at $\r=\r_K$, \
$\r_K=\sqrt{2}\r_h > \r_h$.

 One may  expect  that
 this  naked  singularity
  may be
resolved  by the enhan\c{c}on mechanism
\cite{enh1} (as was originally expected \ci{kts} 
for the singularity
in the conifold case). First, the
singularity is of the right  repulson type
\cite{rep1}. Second important feature
is  the underlying $SU(2)$
symmetry of the 6-d part of the metric (see 
 footnote 2 in \cite{enh1}).
To  make the argument for such  resolution
 at a quantitative level
 is, however,  non-trivial.\foot{Ref. \cite{enh1} used the   
 form of the
effective action for  the  D6-branes wrapped over
 on K3  that probe the geometry.
The case of D6 on K3  is
similar to the case of D4 on K3 which has been  extensively  discussed
in
the literature \cite{ea1}.
In the present case
 we have  a Calabi-Yau space  of dimension 6
 and the
D5-brane  we are dealing with 
 here is wrapping a two-cycle rather than 
the whole
space. Thus we have nontrivial tangent and normal
 bundles which will
affect  the Chern-Simons term.
The  Dirac-Born-Infeld part of the action is also
different (see \cite{ea4} for details).}

If a mechanism  similar to the one in   \cite{enh1}
 does apply in the present case,
 then the
 geometry  should ``stop"  at
$\r=\r_K  $ before reaching the singularity at $\r_h$.
Expanding around $\r=\r_K  $
we get
\be
K(\r_K + \tilde \r)= {2Q\over \r_K} \tilde \r+ {O}(\tilde \r^2)\ ,
 \ \ \ \ \ \ \
h(\r_K + \tilde \r)= h_0+ {Q^2\over 2g_sP^2a^4} + {O}(\tilde \r^2)\ .
\ee
This  is similar to the IR behavior found
in the deformed conifold case  \cite{ks}.
In particular,  the constant value of the
 warp factor $h$ at short distances  should imply again
  confinement in the IR.


\section{Superpotential and first order system}


Let us now  demonstrate how the 1-st order system of equations
and the solution of the previous section can
be derived directly without using the expression
for the resolved conifold metric.
 We shall  follow  the  original 
  approach of \cite{kts},
 i.e. start with an ansatz for the 10-d metric and p-form fields
 which has the required  symmetries, compute the 1-d action
for the radial evolution
 that reproduces  the  type IIB  supergravity equations of motion
 restricted to this ansatz,  show that this action admits a 
  superpotential  and thus obtain  a 1-st order system.
 
As we  shall explain,  the  same strategy applies also 
 to  the case 
 of the deformed conifold  ansatz
  considered in \ci{ks}. The corresponding  superpotential 
  has essentially the same structure as in the conifold \ci{kts} and 
  resolved conifold case, and reproduces 
  the 1-st order  system found in \ci{ks} 
  thus checking its consistency.
  
\subsection{Resolved conifold case}

Let us   choose  the
10-d metric in the following  ``5+5" form
\be
ds^2_{10}  =  e^{2 p- x } (e^{2A} dx^\m dx^\m + du^2 )  +
 \left[ e^{-6p  - x} e_{\psi}^2  +  e^{x+y}  \big(
e_{\theta_1}^2 + e_{\phi_1}^2\big)
+   e^{x-y}    \big( e_{\theta_2}^2 + e_{\phi_2}^2\big)\right] \ ,
\la{rerr}
\ee
where $A,p,x,y$ are  functions  of a radial coordinate $u$. 
Note that the metric of the previous section 
 \rf{mee},\rf{reso}
belongs to this class  ($u$ is related to $\r$).
To be able to describe the resolved conifold
case we have included the function $y$ which measures an ``asymmetry"
between  the two $S^2$ parts  ($y$ was set to zero in the ``symmetric" 
conifold  ansatz \ci{kts}).
The ansatz for the remaining fields will be the same
as in \rf{fields},\rf{rrr},\rf{fre}, i.e.\foot{In this section prime will denote 
derivatives over $u$.}
 \be
H_3=du\wedge \big[f'_1(u)e_{\theta_1}\wedge
e_{\phi_1}+f'_2(u) e_{\theta_2}\wedge e_{\phi_2}\big] \ , \la{vvo}
\ee
\be
F_3=P e_{\psi}\wedge \big( e_{\theta_2}\wedge
e_{\phi_2}-e_{\theta_1}\wedge e_{\phi_1}\big)\ , \la{vot}
\ee
\be
F_5={\cal F} +*{\cal F}\ , \ \  \ \ \ \ \ \
 {\cal F}= K(u) e_{\psi}\wedge
e_{\theta_1}\wedge e_{\phi_1}\wedge e_{\theta_2}\wedge e_{\phi_2}\ ,
\la{lop} \ee
\be
  K(u) \equiv Q + P [f_1(u) - f_2(u)] \  . \la{flop}
\ee
Here we have    explicitly used the constraint \rf{kkk}
following from the Bianchi identity for the 5-form field
(so that the Bianchi identities for all three p-form fields 
are satisfied).
Thus  only $f_1$ and $f_2$ will be considered  as 
independent functions of $u$ coming out of 
the p-form part \rf{vvo}--\rf{flop}  of the
ansatz.
We shall assume that the axion is zero (this  is consistent with
\rf{vot}) but will   keep the dilaton $\P=\P(u)$.

The type IIB supergravity equations of motion follow from the action
$$
 S_{ 10} =-{1\over 2\kappa_{10}^2}  \int d^{10} x \bigg( \sqrt{-g_{10}}
\bigg[ \ R_{10}
 - { \textstyle{1\over 2}} (\partial \Phi)^2
- { \textstyle{1\over 12}} e^{-\Phi}   (\partial B_2)^2  $$
 \be - \ { \textstyle{1\over 2}}  e^{2 \Phi} (\partial {\cal C})^2
 - { \textstyle{1\over 12}} e^{ \Phi}  (\partial C_2  - {\cal C}
\partial B_2) ^2
- { \textstyle{1 \over 4\cdot 5!}}  F^2_5\ \bigg]
- { \textstyle{1\over 2\cdot 4! \cdot (3!)^2    }}
 {\ep_{10}} C_4 \partial C_2 \partial B_2 + ... \bigg) \ , \la{acto}
 \ee
 $$(\partial B_2)_{...} =3 \partial_{[.} B_{..]} ,\ \  \ 
\   (\partial C_4)_{....} \equiv 5 
\partial_{[.} C_{....]}  , \ \
F_5= \partial C_4 + {5} (B_2 \partial
 C_2 - C_2 \partial B_2) , $$
supplemented with the on-shell constraint $F_5=*F_5$ \ci{iib}.
The  1-d action reproducing
the resulting equations of motion restricted to the above ansatz
has the following general structure
\begin{equation}
 S
= c  \int du
\ e^{4 A} \bigg[ 3 A'^2
- \ha G_{ab}(\varphi)  \varphi'^a  \varphi'^b -
 V(\varphi)\bigg]
 \ ,  \la{vvv}  \ee
 where $c = -4  {Vol_9\over 2\kappa_{10}^2}$.
 It should be supplemented with   the ``zero-energy" constraint
 \be
 3 A'^2
- \ha G_{ab}(\varphi)  \varphi'^a  \varphi'^b +
 V(\varphi)= 0 \ . \la{ret}
 \ee
 The existence of a superpotential (usually associated with
residual supersymmetry,  see, e.g.,   \cite{rg1,rg2}
but also \ci{rg3})
 means that $V$ in \rf{vvv}
 can be represented in the form
\begin{equation}\la{super}
V ={\tx  {1\over 8}} G^{ab} {\partial W\over \partial \varphi^a}
{\partial W\over \partial \varphi^b} - {\tx {1\over 3}} W^2
\ .
\end{equation}
In this case the 2-nd order equations following from \rf{vvv}
and  the constraint \rf{ret}
are satisfied on the solutions of the 1-st order system
 \begin{equation}
\varphi'^a = { \ha } G^{ab} { \partial  W
 \over \partial \varphi^b}
\ ,
\ \ \ \ \ \ \ \
A' = - {\tx { 1 \over 3}} W (\varphi) \ .\la{flo}
\end{equation}
In  our present case  we have 6 dynamical variables
$\varphi^a=(x,y,p,\Phi,f_1,f_2)$.
As follows from  \rf{acto} in the case of the ansatz
\rf{rerr}--\rf{lop} 
\be
G_{ab}(\varphi)  \varphi'^a  \varphi'^b
= x'^2+ \ha y'^2+6p'^2 +  \fo \bigg[
\Phi'^2  +   P^2 e^{-\Phi- 2x}
 ( e^{ - 2y}  f'^2_1 + e^{ 2y} f'^2_2) \bigg] \ , \la{kin}
 \ee
 \be
V(\varphi) = \fo e^{-4p-4x}\cosh 2y - e^{2p-2x}\cosh y+
 {\tx {1\over 8}} e^{8p}  \bigg(  2 P^2 e^{\Phi-2x}\cosh 2y
+   e^{-4x} K^2 \bigg)\ , \la{pot}
\ee
where we  separated the 
gravity  contributions 
 (coming from the $R_{10}$-term in \rf{acto})
 from the ``matter" ones 
and it is  assumed that $K$ is a combination of $f_1,f_2$ in \rf{flop}.

Similar expressions  corresponding to the case of
\be y=0\ , \ \ \ \ \ \ \ \ \ f'_1=-f'_2 \la{spe}
 \ee
  appeared  in  \ci{kts}.
Indeed, that  restriction was  consistent. 
As follows from \rf{vvv} and \rf{kin},\rf{pot} the equation
for $y$ is satisfied  by $y=0$  if $f'^2_1=f'^2_2$.
Also,  the potential  \rf{pot}
depends only on one of the two combinations
 $f_\pm \equiv f_1\pm f_2$, so that  the equation for $f_+$ is satisfied
 automatically by $f_+'=0$ if  $y=0$.
 The 1-d action of \ci{kts} may be found  by  
 eliminating  $f_+$ from the action using its 
 equation of motion and then setting $y=0$.

It is quite remarkable that, just like  in  the ``symmetric"  case
considered in \ci{kts},  the more general system \rf{kin},\rf{pot}
still admits a  simple superpotential $W$ given
  by the direct superposition
of the  gravitational and  matter parts 
$$  
W(\varphi)
=e^{4p}+e^{-2p-2x}\cosh y  + \ha  e^{4p-2x} K $$
\be  = \ e^{4p}+e^{-2p-2x}\cosh y  + \ha  e^{4p-2x} [Q +
P(f_1-f_2)]\ . 
\la{wew}
\end{equation}
Note that  the dilaton
factors in the kinetic \rf{kin} and  potential \rf{pot}
terms conspire so that
 the  superpotential  does not depend on the dilaton. This 
implies  that $\P=\const$ on  the solution of the resulting 1-st order 
system of equations 
 \rf{flo} for $A,x,y,p,\P, f_1,f_2$
\be
x'= - e^{-2p-2x}\cosh y   - \ha  e^{4p - 2x} K  \ , \ \ \ \ \ \ \ \ \ \
y' = e^{-2p - 2 x } \sinh y\ ,
\la{sy} \ee
\be\la{syy}
p'=  {\tx {1\over 3}}
e^{4p} - \si  e^{-2p-2x} \cosh y + \si e^{4p-2x} K \ , \
\ee
\be\la{syyy} 
A'=-{\tx{1\over 3}}
 e^{4p} - {\tx{1\over 3}} e^{-2p-2x} \cosh y
   -   \si   e^{4p - 2x } K \ ,
\ee
\be
 f'_1 =  P e^{\Phi + 4p +2y }  \ , \ \ \ \ \  \ 
f'_2 = - Pe^{\Phi + 4p - 2y }\ ,\ \ \  \ \ \ \ \P'=0 \ .  \la{sys}
\ee
We see that \rf{spe}   corresponding to the ``standard"
conifold case is indeed a special solution of this more general 
system.

To establish the equivalence
of this system with the one found in the previous section,
it is useful first 
to look at  the ``gravitational sector" equations
 that do not depend on matter functions $f_i$.
 Since the superpotential \rf{kin} 
 is the direct sum of the gravitational and matter 
 terms, $M^{10}=R^4 \times$(resolved conifold)
  \ should be  a solution to these
 equations with  $K=0$. 
Indeed, as  follows from \rf{sy}--\rf{syyy}
 the factor $e^{2p-x+2A}$ that multiplies $R^4$ part in 
 \rf{rerr} satisfies
\be\la{koo}
h'=-  K h  e^{ 4p -2x}\ ,
\ \ \ \ \ \ \ \ \  
h^{-1/2} \equiv e^{2p-x+2A}\ ,
\ee
so that  $h$ (which at the end 
should be the same as in \rf{mee}) is 
constant if $K$ is set equal
to 0. The equations for $x,y,p$  with $K=0$ imply 
\be
{dx\ov dy} = - \coth y  \ , \ \ \ \ \ \ \  e^{2x} = b^2 \sinh^{-2} y
\ , \la{syk} \ee
\be\la{syyk}
 {dq\ov dy} =  2 \b^3  (\sinh y)^{-4}  e^{q}   \ , \      
 \ \ \ \ \  \     e^{-q}  =  \b^3 \bigg( {\cosh y - { 1 \ov 3 } \cosh 3y \ov
 \sinh^{3} y }  - c \bigg)\ , \ \ \ \ \       q\equiv 6p -x    \ . 
\la{yut} \ee
For a special  choice of the integration constants
$\b =-\ha a^2, \ c={ 4 \ov 3}  $  
we  finally 
 reproduce
 (using $dy= e^{-2p-2x} \sinh y\ du$  and introducing  $\rho$
 instead of $u$ to get  simple analytic expressions)
  the resolved conifold metric \rf{reso} (cf. \rf{rerr})
\be
e^{2y} = {\r^2 \ov \r^2 + 6 a^2} \ , \ \ \ \ \ \ 
e^{2x} = {\tx { 1 \ov 36} } \r^2 (\r^2 + 6 a^2) \ , \ \ \ \ \ \ 
e^{-6p+x}=  {\tx { 1 \ov 324} } \r^4 (\r^2 + 9 a^2) \ .\la{ress}
\ee
The resolution parameter $a$ is thus one of the three integration
 constants in the above 1-st order system.
 
In the presence of matter the 
 full system \rf{sy}--\rf{sys}
 may be solved by first concentrating  on the equations that do not 
involve $K$, i.e. on the equation for $y$ 
and for $z=x+3p$. It is useful to introduce 
the new radial direction $t$, \  $dt=e^{-2p - 2 x}  du$
so that they become $ {dy\ov dt} =   \sinh y , \ 
{dz\ov dt} = - e^{2z} +{\tx { 3\ov 2} }\cosh y $.
One then finds that the 
ratios of the first and 
the second,  and the second and the third
coefficients in the ``internal"  5-d  part of the metric 
\rf{rerr}, i.e. $ e^{-6p-x}/e^{x+y}  = e^{-2z-y}$ and 
$e^{x+y}/e^{x-y}= e^{2y}$,  are the same as in the resolved 
conifold metric, in agreement with \rf{mee}.
The rest of the equations then
become equivalent (for the special choice of the integration constants)
to the
system in the previous section.

\subsection{Deformed  conifold  case}

Let us now  complement  the discussion of the
deformed conifold case in \ci{ks} by demonstrating explicitly that
the first order-system  there  also follows
from a simple superpotential
which has essentially the same structure as \rf{wew}.

Motivated by the form  of the deformed conifold metric 
\rf{ddd},\rf{gee} let us make  the following  ansatz for the metric
(cf. \rf{rerr}) 
\be\la{dddf}
ds^2 =  e^{2 p-x} (e^{2A}  dx^\mu dx^\mu + du^2) 
  + 
\left[ e^{-6p - x} \gp^2  +   e^{x+y}    ( \go^2 + \gd^2)
+  e^{x-y}  ( \gt^2 + \gc^2)
 \right] \ .  \ee
The ansatz for the p-forms is the same as in \ci{ks}
(cf. \rf{vvo}--\rf{flop})
\be
H_3  = du\wedge[  f'(u) \go \we \gd + k'(u) \gt \we \gc] \ , \ \ \
\la{he} \ee
\be 
F_3 = F(u) \go\we\gd\we\gp + [2P-F(u)] \gt\we\gc\we\gp +
F'(u) du \we ( \go \we \gt +  \gd \we \gc)
\ , \la{hhe}
\ee
\be
 F_5 = \F_5 + \F_5^* \ , \ \ \ \ \ \ \
 \F_5 =    K(u) \go\we\gd\we\gt \we\gc\we\gp\ , \la{poi}
 \ee
 \be 
 K(u)  \equiv Q +  k(u)  F(u) + f(u) [2P-F(u)] \ , \la{uui}
 \ee
 where $F,f,k$ are functions to be determined and $P$ and $Q$ are
 constants.
As in the previous case \rf{vvo},\rf{vot},\rf{lop}, we 
explicitly ensure that 
the Bianchi identities for the p-forms are satisfied automatically. 
The independent functions of $u$ which will appear in the 
 1-d action \rf{vvv} are thus $A$ and 
$\varphi^a=(x,y,p,\Phi,f,k,F)$. 
The corresponding  kinetic and potential terms 
in \rf{vvv} are found to be similar to \rf{kin},\rf{pot}
\be
G_{ab}(\varphi)  \varphi'^a  \varphi'^b
= x'^2+ \ha y'^2+6p'^2 +  \fo \bigg[
\Phi'^2  +   e^{-\Phi- 2x}
 ( e^{ - 2y}  f'^2 + e^{ 2y} k'^2) + 2  e^{\Phi- 2x} F'^2  \bigg] \ , 
 \la{kine}
 \ee
 $$
V(\varphi) = \fo e^{-4p-4x}- e^{2p-2x}\cosh y 
 + \fo e^{8p}\sinh^2 y  $$ \be 
 +\ 
 {\tx {1\over 8}} e^{8p}  \bigg[ \ha  e^{-\Phi-2x} (f-k)^2 
 +  e^{\Phi-2x}  [ e^{  -2y}  F^2  + e^{ 2y} (2P-F)^2]
+  e^{-4x} K^2  \bigg]\ ,  \la{pote}
\ee
where $K$ is the combination of the independent functions
$f,k,F$  given in   \rf{uui}.
The corresponding 
 superpotential satisfying \rf{super}
  again  does not depend on the dilaton 
 and 
  is a sum of the gravitational and matter parts, i.e.  
   has essentially the same structure as 
  the previous one \rf{wew}
  $$
  W(\varphi) =    e^{4p} \cosh y + e^{-2 p-2x}
   +   \ha  e^{4p - 2x } K    $$
   \be = \  e^{4p} \cosh y + e^{-2 p-2x}
   +   \ha  e^{4p - 2x } [ Q +  k  F+ f(2P-F)]
    \ .  \la{wewa}
   \ee
Thus  there is a close similarity (``duality") 
 between  the  1-st order systems for the  ``resolved" and ``deformed"  
cases.
 
{}From \rf{flo} and \rf{wewa} we find 
  the following set of 1-st order equations for the independent functions  
 $A,x,y,p, f,k,F,\P$
 \be
x'= - e^{-2p-2x}    - \ha  e^{4p - 2x} K  \ , \ \ \ \ \ \ \ \ \ \
y' = e^{4p  } \sinh y\ ,
\la{syd} \ee
\be\la{syyd}
p'=  {\tx {1\over 3}}
e^{4p}\cosh  y  - \si  e^{-2p-2x}  + \si e^{4p-2x} K \ , \
\ee
\be\la{syyyd} 
A'=-{\tx{1\over 3}}
 e^{4p}\cosh y  - {\tx{1\over 3}} e^{-2p-2x} 
   -   \si   e^{4p - 2x } K \ ,
\ee
\be
 f' =    e^{\Phi + 4p +2y } (2P-F)  \ , \ \ \ \
k' =   e^{\Phi + 4p - 2y }F \ ,\ \ \  
F' = - \ha e^{-\Phi + 4p } (f-k)\ ,   \ \ \ \ \P'=0 \ .  \la{sysd}
\ee
The  special solution is 
$y=0, \ f=k\ , F=  P$. 
In this case the  system
 \rf{syd}--\rf{sysd} becomes identical
 to \rf{sys}--\rf{sys} with $f_1=-f_2=f$ and both 
 reduce to  the ``standard"  conifold case  of \ci{kts}. 
  
To show that this system contains  the 
solution of \ci{ks} we  follow the 
same strategy as in the previous subsection: first analyze 
the subset of  gravitational sector equations to find  that 
the ratios of the functions in the metric  are the same as in the
case of the deformed conifold
and then include the matter part.
We again find the relation \rf{koo} implying  4+6 
factorization of the metric  for $K=0$. 
 The equations for $x,y,p$  with $K=0$ here lead to 
 the relations which are very similar  (``dual")
  to \rf{syk},\rf{yut}, 
    $
 {dq\ov dy} =  2 \coth y   ,\  e^q = { \b^2 \sinh^2 y}, \
   q\equiv 6p -x ,$ and 
 $
{dx\ov dy} = - e^{-q} e^{-3x} (\sinh y )^{-1}  = - e^{-3x}  \b^{-2}
(\sinh y )^{-3}$.
It is useful 
to introduce  the new radial coordinate $\ta$ 
so that 
\be
{ dy \ov d \ta} = - \sinh y \ , \ \ \ \ \ \ \ \ e^{y} =  \tanh \rt  \ ,
\  \ \ \ \ \ \ \  d\ta \equiv  - e^{4p} du\
, 
 \la{qrr}
 \ee
\be
{ d(x-6p)\ov d \ta}    =  2 \cosh y \ , \ \ \ \ \ \ \ \ \ \ \ \ 
e^{x-6p} = \b^2  \sinh^2\ta \ , 
\la{uoi}
\ee
where $b$ is an  integration constant.
The remaining equation  is
\be
{ dx \ov d \ta} =  e^{-6p-2x} =  e^{x-6p} e^{-3x} 
=  \b^2  \sinh^2\ta\   e^{-3x} \ , 
\la{oop}
\ee\be
e^{3x}   = c   +  {\tx {3  \ov 2}} \b^2 (\ha \sinh 2\ta - \ta)    
\ . \la{uop}
\ee
Choosing $c=0, \ \b^2 = { 1 \ov 96} \ep^4$   we thus reproduce
the deformed conifold metric \rf{ddd}
\be
e^{-6p - x} = e^{2p-x} e^{-8p} = \si \ep^{4/3} \K^{-2} \ , \ \  \ \ \
e^{x+y} = \ha \ep^{4/3} \K  \sinh^2 \rt \ ,  \ \ \
e^{x-y} = \ha \ep^{4/3} \K  \cosh^2 \rt \ ,  \la{fina}
\ee
where $\K(\tau)$ was  defined in \rf{kee}
and $\ep$ is the deformation parameter.

It is quite remarkable that making simple ansatze \rf{rerr} or
\rf{dddf} for the 6-d part of the metric  one finds that the 
1-d action leading to the associated
Einstein (Ricci-flatness) equations admit a superpotential, 
and that  the solutions of the corresponding 1-st order systems
are the resolved \rf{reso} and the deformed \rf{ddd} conifold metrics
respectively!

In the general case 
the system \rf{syd}--\rf{sysd}
can be solved by starting with the equations that do not involve matter
functions:
equation for  $y$ \rf{qrr}
and the following  combination of the equations for $x$ and $p$ 
\be
{d(3p+x) \ov d\ta} =- \cosh y    + {\tx { 3\ov 2} } e^{-2(3p+x)}
     \ . \la{kyo}
  \ee
  This equation is  solved   by first introducing 
 $ w= 3p+x  +  \ln \sinh \ta $. As a result, 
 \be
 e^{3p+x} = \sqrt{ {\tx{3\ov 2}}} (\sinh\ta )^{-1}
       (\ha \sinh 2\ta - \ta)^{1/2} \ ,  
 \ee
 where we set the integration constant  to zero so 
 that $e^{3p+x}$ is exactly the same as in  \rf{fina}. 
 
 This implies that as in the resolved conifold 
 case of the previous subsection, 
 the ratios of the coefficients in the internal 5-d part of the metric 
 \rf{dddf}, i.e. $ e^{-6p-x}/e^{x+y}  = e^{-2(3p+x)-y}$ and 
$e^{x+y}/e^{x-y}= e^{2y}$, 
 have the same values as in the 
deformed conifold  metric \rf{ddd}.
The solution 
of the full system is then equivalent to  that of \ci{ks}
for the ``D3-brane"  ansatz \rf{mee}.

\section*{Acknowledgments}
We are  grateful to S. Frolov, I. Klebanov and G. Papadopoulos for many
useful
discussions and comments.
LAPZ would
like to acknowledge the Office of the Provost at the University of
Michigan and the High Energy Physics Division of the Department of
Energy for support. The work of AAT  was  supported in part by
the DOE grant DE-FG02-91R-40690,
INTAS project 991590
and PPARC SPG grant  PPA/G/S/1998/00613.
We would like also to thank the  Aspen Center for Physics
for hospitality while
 part of this work was done.



\begin{thebibliography}{99}
\bibitem{pol}
A.M.~Polyakov,
``String theory as a universal language,''
hep-th/0006132.
``The wall of the cave,''
Int.\ J.\ Mod.\ Phys.\  {\bf A14} (1999) 645
[hep-th/9809057].
``String theory and quark confinement,''
Nucl.\ Phys.\ Proc.\ Suppl.\  {\bf 68} (1998) 1
[hep-th/9711002].




\bibitem{ads1}
J.~Maldacena,
``The large N limit of superconformal field theories and
supergravity,''
Adv.\ Theor.\ Math.\ Phys.\  {\bf 2} (1998) 231
[hep-th/9711200].
S.S.~Gubser, I.R.~Klebanov and A.M.~Polyakov,
``Gauge theory correlators from non-critical string theory,''
Phys.\ Lett.\  {\bf B428} (1998) 105
[hep-th/9802109].
E.~Witten,
``Anti-de Sitter space and holography,''
Adv.\ Theor.\ Math.\ Phys.\  {\bf 2} (1998) 253
[hep-th/9802150].



\bi{moo}
M.R.~Douglas and G.~Moore,
``D-branes, Quivers, and ALE Instantons,''
hep-th/9603167.





\bibitem{kehagias}
A.~Kehagias,
``New type IIB vacua and their F-theory interpretation,''
Phys.\ Lett.\  {\bf B435} (1998) 337
[hep-th/9805131].


\bibitem{kw1}
I.R.~Klebanov and E.~Witten,
``Superconformal field theory on threebranes at a Calabi-Yau
singularity,''
Nucl.\ Phys.\  {\bf B536} (1998) 199,
[hep-th/9807080].

\bibitem{mr}
D.R.~Morrison and M.R.~Plesser,
``Non-spherical horizons. I,''
Adv.\ Theor.\ Math.\ Phys.\  {\bf 3} (1999) 1
[hep-th/9810201].

\bibitem{kg}
S.S.~Gubser and I.R.~Klebanov,
``Baryons and domain walls in an N = 1 superconformal gauge
theory,''
Phys.\ Rev.\  {\bf D58} (1998) 125025
[hep-th/9808075].


\bibitem{kn}
I.R.~Klebanov and N.A.~Nekrasov,
``Gravity duals of fractional branes and logarithmic RG flow,''
Nucl.\ Phys.\  {\bf B574} (2000) 263
[hep-th/9911096].


\bibitem{gim}
E.G.~Gimon and J.~Polchinski,
``Consistency Conditions for Orientifolds and D-Manifolds,''
Phys.\ Rev.\  {\bf D54} (1996) 1667
[hep-th/9601038]. 
M.R.~Douglas,
``Enhanced gauge symmetry in M(atrix) theory,''
JHEP {\bf 9707} (1997) 004
[hep-th/9612126].


\bibitem{kts}
I.R.~Klebanov and A.A.~Tseytlin,
``Gravity duals of supersymmetric $SU(N) \times  SU(N+M)$ gauge theories,''
Nucl.\ Phys.\  {\bf B578} (2000) 123
[hep-th/0002159].


\bibitem{ks}
I.R.~Klebanov and M.J.~Strassler,
``Supergravity and a confining gauge theory: Duality cascades and
(chi)SB-resolution of naked singularities,''
hep-th/0007191.

\bibitem{vaf1}
R.~Gopakumar and C.~Vafa,
``On the gauge theory/geometry correspondence,''
hep-th/9811131.

\bibitem{kw2}
I.R.~Klebanov and E.~Witten,
``AdS/CFT correspondence and symmetry breaking,''
Nucl.\ Phys.\  {\bf B556} (1999) 89
[hep-th/9905104].

\bibitem{vaf2}
C.~Vafa,
``Superstrings and Topological Strings at Large N,''
hep-th/0008142.


\bibitem{MN}
J.M.~Maldacena and C.~Nu\~nez,
``Towards the large n limit of pure N = 1 super Yang Mills,''
hep-th/0008001.


\bibitem{ig}
I.R.  Klebanov, private communication.

\bibitem{polg}
M.~Gra\~na and J.~Polchinski,
``Supersymmetric three-form flux perturbations on AdS(5),''
hep-th/0009211.

\bibitem{gub}
S.S.~Gubser,
``Supersymmetry and F-theory realization of the deformed conifold
with
three-form flux,''
hep-th/0010010.



\bibitem{candelas}
P.~Candelas and X.C.~de la Ossa,
``Comments On Conifolds,''
Nucl.\ Phys.\ {\bf B342} (1990) 246.


\bibitem{horst}
G.T.~Horowitz and A.~Strominger,
``Black strings and P-branes,''
Nucl.\ Phys.\  {\bf B360} (1991) 197.

\bibitem{dul}
M.J.~Duff and J.X.~Lu,
``The Selfdual type IIB superthreebrane,''
Phys.\ Lett.\  {\bf B273} (1991) 409.


\bibitem{enh1}
C.V.~Johnson, A.W.~Peet and J.~Polchinski,
``Gauge theory and the excision of repulson singularities,''
Phys.\ Rev.\  {\bf D61} (2000) 086001
[hep-th/9911161].


\bibitem{page}
D.N.~Page and C.N.~Pope,
``Which Compactifications Of D = 11 Supergravity Are Stable?,''
Phys.\ Lett.\  {\bf B144} (1984) 346.



\bibitem{minasian}
R.~Minasian and D.~Tsimpis,
``On the geometry of non-trivially embedded branes,''
Nucl.\ Phys.\  {\bf B572} (2000) 499
[hep-th/9911042].

\bibitem{papa}
G.~Papadopoulos, J.G.~Russo and A.A.~Tseytlin,
``Curved branes from string dualities,''
Class.\ Quant.\ Grav.\  {\bf 17} (2000) 1713
[hep-th/9911253].


\bibitem{ohta}
K.~Ohta and T.~Yokono,
``Deformation of conifold and intersecting branes,''
JHEP {\bf 0002} (2000) 023
[hep-th/9912266].




\bibitem{rep1}
K.~Behrndt,
``About a class of exact string backgrounds,''
Nucl.\ Phys.\  {\bf B455} (1995) 188
[hep-th/9506106].
R.~Kallosh and A.~Linde,
``Exact supersymmetric massive and massless white holes,''
Phys.\ Rev.\  {\bf D52} (1995) 7137
[hep-th/9507022].
M.~Cvetic and D.~Youm,
``Singular BPS saturated states and enhanced symmetries of
four-dimensional N=4 supersymmetric string vacua,''
Phys.\ Lett.\  {\bf B359} (1995) 87
[hep-th/9507160].


\bibitem{ea1}
M.~Bershadsky, C.~Vafa and V.~Sadov,
``D-Strings on D-Manifolds,''
Nucl.\ Phys.\  {\bf B463} (1996) 398
[hep-th/9510225].
M.B.~Green, J.~A.~Harvey and G.~Moore,
``I-brane inflow and anomalous couplings on D-branes,''
Class.\ Quant.\ Grav.\  {\bf 14} (1997) 47
[hep-th/9605033].
K.~Dasgupta, D.P.~Jatkar and S.~Mukhi,
``Gravitational couplings and Z(2) orientifolds,''
Nucl.\ Phys.\  {\bf B523} (1998) 465
[hep-th/9707224].

\bi{ea4}
C.P.~Bachas, P.~Bain and M.B.~Green,
``Curvature terms in D-brane actions and their M-theory origin,''
JHEP {\bf 9905} (1999) 011
[hep-th/9903210].





\bibitem{rg1}
D.Z.~Freedman, S.S.~Gubser, K.~Pilch and N.P.~Warner,
``Renormalization group flows from holography supersymmetry and a
c-theorem,''
hep-th/9904017.

\bibitem{rg2}
L.~Girardello, M.~Petrini, M.~Porrati and A.~Zaffaroni,
``The supergravity dual of N = 1 super Yang-Mills theory,''
Nucl.\ Phys.\  {\bf B569} (2000) 451
[hep-th/9909047].

\bi{rg3}
O.~DeWolfe, D.Z.~Freedman, S.S.~Gubser and A.~Karch,
``Modeling the fifth dimension with scalars and gravity,''
Phys.\ Rev.\  {\bf D62}, 046008 (2000)
[hep-th/9909134].
K.~Skenderis and P.K.~Townsend,
``Gravitational stability and renormalization-group flow,''
Phys.\ Lett.\  {\bf B468}, 46 (1999)
[hep-th/9909070].
K.~Behrndt and M.~Cvetic,
``Supersymmetric domain wall world from D = 5 simple gauged
supergravity,''
Phys.\ Lett.\  {\bf B475}, 253 (2000)
[hep-th/9909058].



\bibitem{iib}
J.H.~Schwarz,
``Covariant Field Equations Of Chiral N=2 D = 10 Supergravity,''
Nucl.\ Phys.\  {\bf B226}, 269 (1983).
E.~Bergshoeff, C.~Hull and T.~Ortin,
``Duality in the type II superstring effective action,''
[hep-th/9504081].

\end{thebibliography}
\end{document}